# Material-specific high-resolution table-top extreme ultraviolet microscopy

*Wilhelm Eschen[1,2], Lars Loetgering[1,2,3], Vittoria Schuster[1,2], Robert Klas[1,2], Alexander Kirsche[1,2], Lutz Berthold[5], Michael Steinert[1], Thomas Pertsch[1,4], Herbert Gross[1], Michael Krause[5], Jens Limpert[1,2,4] & Jan Rothhardt [1,2,4]*

*Corresponding author, E-mail: wilhelm.eschen@uni-jena.de*

1. Institute of Applied Physics, Abbe Center of Photonics, Friedrich-Schiller-Universität Jena, Albert-Einstein-Str. 15, 07745 Jena, Germany
2. Helmholtz-Institute Jena, Fröbelstieg 3, 07743 Jena, Germany
3. Leibniz Institute of Photonic Technology, Albert-Einstein-Straße 9, 07745, Jena, Germany
4. Fraunhofer Institute for Applied Optics and Precision Engineering IOF, Albert-Einstein-Str. 7, 07745 Jena, Germany
5. Fraunhofer Institute for Microstructure of Materials and Systems IMWS, Walter-Hülse-Str. 1, 06120 Halle, Germany

**Abstract:**

Microscopy with extreme ultraviolet (EUV) radiation holds promise for high-resolution imaging with excellent material contrast, due to the short wavelength and numerous element-specific absorption edges available in this spectral range. At the same time, EUV radiation has significantly larger penetration depths than electrons. It thus enables a nano-scale view into complex three-dimensional structures that are important for material science, semiconductor metrology, and next-generation nano-devices. Here, we present high-resolution and material-specific microscopy at 13.5 nm wavelength. We combine a highly stable, high photon-flux, table-top EUV source with an interferometrically stabilized ptychography setup. By utilizing structured EUV illumination, we overcome the limitations of conventional EUV focusing optics and demonstrate high-resolution microscopy at a half-pitch lateral resolution of 16 nm. Moreover, we propose mixed-state orthogonal probe relaxation ptychography, enabling robust phase-contrast imaging over wide fields of view and long acquisition times. In this way, the complex transmission of an integrated circuit is precisely reconstructed, allowing for the classification of the material composition of mesoscopic semiconductor systems.

**Introduction:**

Advances in nano-scale metrology of silicon-based devices are crucial for progress in diverse fields, with applications spanning from semiconductor miniaturization[1], energy conversion, and storage, such as next-generation solar cells[2] and battery materials[3], to nanostructures with advanced optical functionality, like meta materials[4] and photonic circuits[5]. The extreme ultraviolet lithography (EUVL) node at 13.5 nm wavelength was selected as a reasoned choice. The electronic structure of tin plasmas provides prominent emission peaks at 13.5 nm wavelength[6,7], while multilayer mirrors such as Mo/Si and Mo/Be reach reflectivities of up to 70%[8,9]. Moreover, at 13.5 nm the penetration depth of EUV photons in silicon is orders of magnitude higher than many elements across the periodic table. Likewise, in comparison to electrons, EUV photons can shed light into the interior of silicon-based environments. The electromagnetic region below the silicon L edge is therefore referred to as the *silicon transparency window* - an ideal testbed for silicon-based substrates and functional materials.

In the past decade, imaging at EUV wavelengths has undergone a transformation. Driven by the increased photon flux and stability available from sources based on high-harmonic generation (HHG)[10–12], lensless imaging techniques have successfully been transferred from large-scale facilities[13], such as synchrotrons and free-electron lasers, to the laboratory[14–16]. In particular, the emergence of ptychography[17,18] has offered a solution to some of the main problems encountered with EUV radiation: its lensless operation principle avoids absorptive losses, aberrations induced by the image forming optics, and its ability to perform wavefront sensing enables the deconvolution of illumination-induced aberrations, resulting in quantitative phase imaging (QPI). These capabilities are afforded by data-driven techniques. A sequence of diffraction patterns is collected on a pixelated detector while the specimen is laterally translated through a focused beam (compare Figure 1). The scan points are chosen in such a way that diffraction patterns from adjacent positions contain overlapping information. In this way, both the illumination wavefront and a phase-sensitive sample micrograph are jointly retrieved. Nevertheless, ptychography with table-top HHG sources[19] is arguably in its infancy. Despite recent highlights in EUV ptychography, such as sub-wavelength resolution on periodic samples[20], bioimaging of hippocampal neurons[21] as well as phase-sensitive reflectometry[22], additional element-specificity is needed to meet the demands from the semiconductor industry and harness the prominent contrast mechanisms in silicon-based environments.

In this work, we present an actively stabilized, high-resolution, quantitative EUV transmission microscope operating in the silicon transparency window. We demonstrate wavelength-scale resolution combined with material-specific imaging. A critical ingredient is the use of structured illumination to overcome the limitations of conventional EUV focusing optics. To this end, an amplitude mask is placed in front of the sample, which shapes the illuminating probe beam. As an algorithmic novelty, we propose orthogonal probe relaxation (OPR)[23] combined with multiple mixed-states (m-s)[24], which results in a precise reconstruction of the specimen's scattering amplitude. Combining the stabilized setup with the structured illumination, we achieve a diffraction-limited half-pitch resolution of 16 nm on a high-contrast, non-periodic sample. In addition, a thin lamella of a silicon-based integrated circuit, extracted from a solid-state disc, is investigated. For the first time, the material composition is retrieved by a single EUV ptychography scan over a wide field of view utilizing the reconstructed scattering amplitude. We believe the current work is an important step forward towards the integration of high-resolution metrology into the realm of EUVL as well as towards material-specific inspection of functional materials in silicon-based environments.

**Results**

*Experimental setup*

Ultrashort fiber laser systems enable average powers in the kilowatt range with millijoule pulse energies[25]. HHG sources operated by these lasers provide high photon flux reaching the milliwatt range for low photon energies[12]. Here, a high-power few-cycle laser is focused into an argon gas jet (see Figure 1). The short (<10 fs) IR driving laser pulses generate a broadband EUV continuum via HHG, which yields a photon flux of $7 \cdot 10^9$ ph/s/eV at 92 eV with particularly high power and pointing stability. More details on the HHG source can be found in the methods section. The EUV radiation is directed and focused onto the sample by means of three multilayer mirrors, which are used to select a central wavelength of 13.5 nm (92 eV) within a 0.2 nm bandwidth (FWHM) window of the broadband continuum. The cascade of three mirrors has been designed to fully compensate for astigmatism induced by the spherical mirrors. The resulting EUV focal spot was characterized by ptychography and exhibits an intensity full-width half-maximum of 3.8 µm and 2.6 µm, which is significantly smaller than in previous two-mirror configurations[26]. More details about the out-of-plane setup and the beam characterization can be found in the supplement. The monochrome EUV beam is directed onto an amplitude mask placed directly in front of the sample. The mask allows for structuring the illumination. Ptychography relies on the precise knowledge of the probe beam with respect to the object at each scan position. Moreover, unaccounted position drifts during the acquisition of a diffraction pattern lead to unwanted decoherence effects[27] and model mismatch[23]. To ensure stable and reliable positioning over hours of measurement time, both the sample and mask holder position are tracked by a laser interferometer and actively stabilized via a feedback loop. Diffraction patterns are recorded by a CCD which is placed 30 mm behind the sample resulting in a detection numerical aperture (NA) of 0.42. Finally, the probe beam and the object are computationally reconstructed by means of ptychographic algorithms[17,18], which are detailed in the methods section.

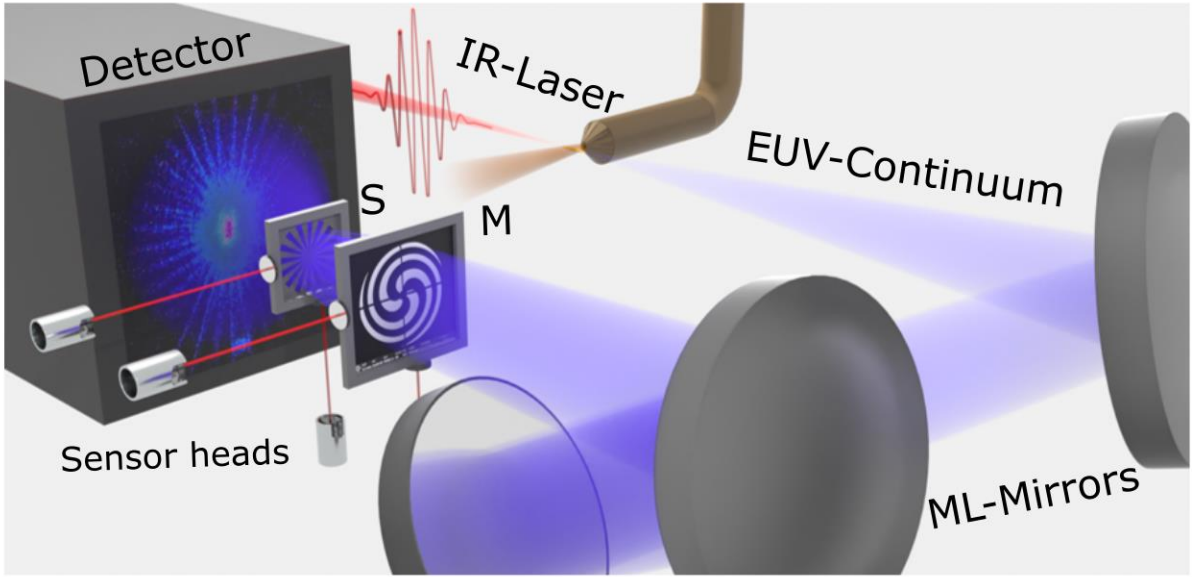

**Figure 1: EUV ptychography setup.** A few-cycle IR-laser is focused in an Argon gas jet where a broad EUV-Continuum is generated. From the broadband continuum, a narrow bandwidth of 0.2 nm is selected at a wavelength of 13.5 nm by three multilayer mirrors (ML-Mirrors) and focused on a mask (M). The sample (S) is illuminated by a structured beam and the resulting diffraction pattern is recorded by the detector.

*EUV-ptychography using structured light*

In recent years, structured illumination has turned out to be beneficial for ptychography for multiple reasons. First, a larger spatial frequency spectrum increases the spread of the diffraction pattern on the detector. The zeroth order of the diffraction is, therefore, less intense and leads to relaxed dynamic range requirements on the detector[28]. Second, the broadened spatial frequency spectrum increases the diffraction-limited resolution[29]. Third, a structured beam improves the convergence of the reconstruction algorithms. Due to the fine structures of the beam a translation of the probe leads to more diverse diffraction patterns, which results in stronger information[30]. So far structured illumination has been used for experiments in the visible[28], soft[31], and hard X-ray[29,30] range. In table-top EUV ptychography multilayer mirrors with relatively low numerical apertures are the prevailing method of choice[19,26,32]. Structured and focused EUV beams have so far only been achieved with a specialized Fresnel zone plate[33], which are however limited in photon efficiency and require additional optical components for spectral selection.

In this work, we shape a structured beam by placing an absorbing nano-structured mask with a fill factor of 50 % at roughly 200 µm upstream of the sample. The combination of multilayer mirrors and the amplitude mask enables a photon efficient spectral selection, structuring, and focusing of the EUV probe beam. Below, we demonstrate that such dedicated illumination engineering pays off by drastically improving the reconstruction quality and lateral resolution.

To investigate the influence of a structured illumination on the EUV image quality, two independent measurements were performed using a smooth (unstructured) and a structured illumination respectively. A pinhole with a diameter of 8 µm is used to restrict the size of the smooth (unstructured) beam. The ptychographic reconstruction of the object is shown in Figure 2 **a.** The colored insets (red, blue) highlight regions where we identified spurious modulations in the reconstructed micrograph (Figure 2 **b**). Clearly, these are imaging artifacts and not real features of the object. The distorted image quality can be attributed to the low translation diversity of the unstructured probe beam[30], which is shown in Figure 2 **d**. The back-propagated complex electric field inside the mask aperture (Figure 2 **e**) matches with the scanning electron microscopy (SEM) image of the fabricated pinhole (small inset). In the mask plane, a slight phase curvate is visible, which is due to the out-of-focus position of the mask.

In the next step, a mask that results in a highly structured beam on the sample is used. The exposure time was doubled to keep the overall number of detected photons per diffraction pattern comparable to the previous experiment using an unstructured pinhole. The resulting reconstruction of the object is displayed in Figure 2 **f**. We observe an overall improved image quality, which is due to a reduction of spurious modulation as compared to the reconstruction obtained from the unstructured beam. By direct comparison of the blue insets in Figure 2 **b** and Figure 2 **g**, we find that a higher resolution is achieved by means of the structured beam. While for the unstructured beam spokes down to a size of 53 nm are resolved, for the highly structured beam features down to a size of 43 nm are resolved (compare enlarged innermost spokes in Figure 2 **c** and Figure 2 **h**). The reconstructed structured beam in the sample plane, which is shown in Figure 2 **i,** was achieved by nanopatterning the mask with a spiral shape[34]. The illumination back-propagated into the mask plane shows the spiral mask (Figure 2 **j**), which matches well with the corresponding SEM images in the inset.

The use of a simple amplitude mask opens up degrees of freedom for functional beam shaping. In addition to enriched spatial frequency content, other properties can be encoded into the illumination wavefront. In this experiment, a charge 3 orbital angular momentum (OAM) was created. Figure 2 **k** indicates an azimuthal lineout of the phase of the reconstructed wavefront along the white, dotted path in Figure 2 **i**, confirming the charge 3 OAM of the beam. Different types of OAM beams can be easily generated with the presented technique by adapting the beam shaping mask[35].

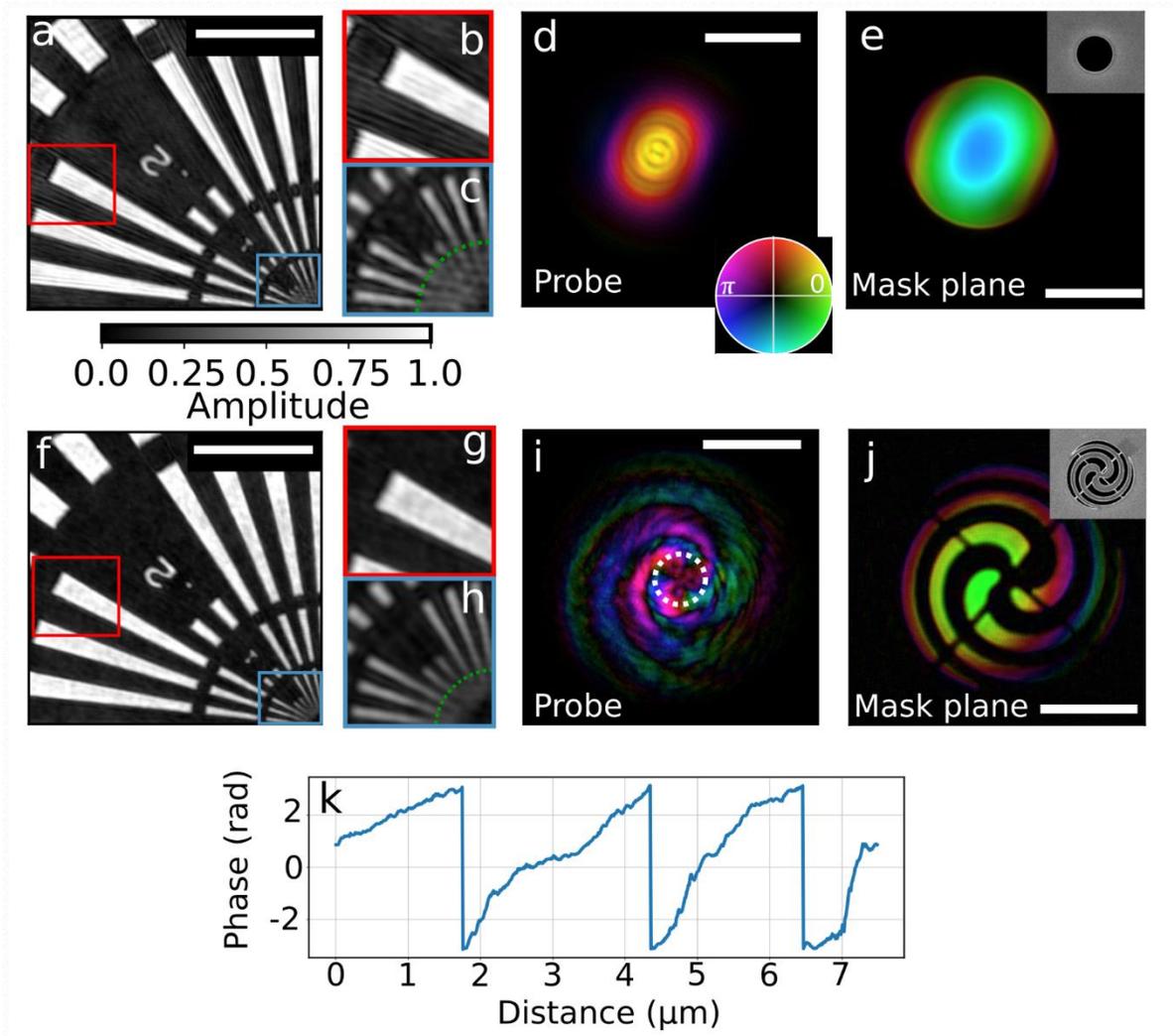

**Figure 2: EUV Ptychography using structured beams. a** Reconstructed transmissivity of the Siemens star using an unstructured illumination. The reconstruction exhibits spurious mid-spatial-frequency modulations, which are displayed in inset **b**. A magnified view of the center of the Siemens star is shown in **c**. The corresponding reconstructed probe is shown in **d**. **e** shows the probe back-propagated into the mask plane. The small inset in **e** shows an SEM image of the mask, which is a 8 µm diameter pinhole. **f** Reconstructed transmission of the Siemens star using structured light showing fewer modulations in **g** and a higher resolution in **h** as compared to the unstructured reconstruction (**b, c**). The green circular line in **c** and **h** corresponds to the smallest radius where the spokes are still resolved. **i, j** Reconstructed illumination in the sample and mask plane. The small inset in **j** shows the SEM image of the mask for comparison. **k** Azimuthal lineout of the probe phase along the white, dotted path indicated in **i**. The scalebar in **a** and **f** indicates 2 µm and the scalebar of **d**, **e**, **i**, **j** corresponds to 5 µm. The brightness and hue of **d**, **e**, **i**, **j** encode modulus and phase, respectively.

### *High-resolution, wide field-of-view ptychographic imaging at 13.5 nm*

The combination of structured illumination and interferometric stability in the positioners allows for high dynamic range (HDR) data acquisition over long-term scans. This enables both wide fields of view and high-resolution imaging. A Siemens star test pattern was scanned for 101 positions. For each position, two diffraction patterns with exposure times of 3 s and 45 s were recorded and stitched to a single high-dynamic-range diffraction pattern. The total acquisition time of the whole scan was 81 minutes. The high-resolution reconstruction of the object is shown in Figure 3 **a**, resulting in a total field of view of 340 µm$^2$. The reconstructed probe is shown in Figure 3 **b**. To demonstrate the flexibility of our mask-based approach, here a charge-1 OAM beam was generated. The achieved high-resolution is shown in the innermost region of the Siemens star, which contains the smallest features and is shown in Figure 3 **c**. The Fourier ring correlation (FRC), a widely used measure for the estimation of resolution in X-ray ptychography[36], was calculated and combined with the half-bit criterion[37]. To this end, a second, independent data set was acquired and the FRC was calculated from both reconstructions, the result of which is shown in Figure 3 d. The FRC curve does not intersect with the half-bit criterion up to the highest detected spatial frequency at the edge of the detector which corresponds to a spatial frequency of 31 µm$^{-1}$ and hence proves a diffraction-limited half-pitch resolution of 16 nm. Note that this resolution analysis is valid over the entire field of view. More details on the reconstruction can be found in the methods section.

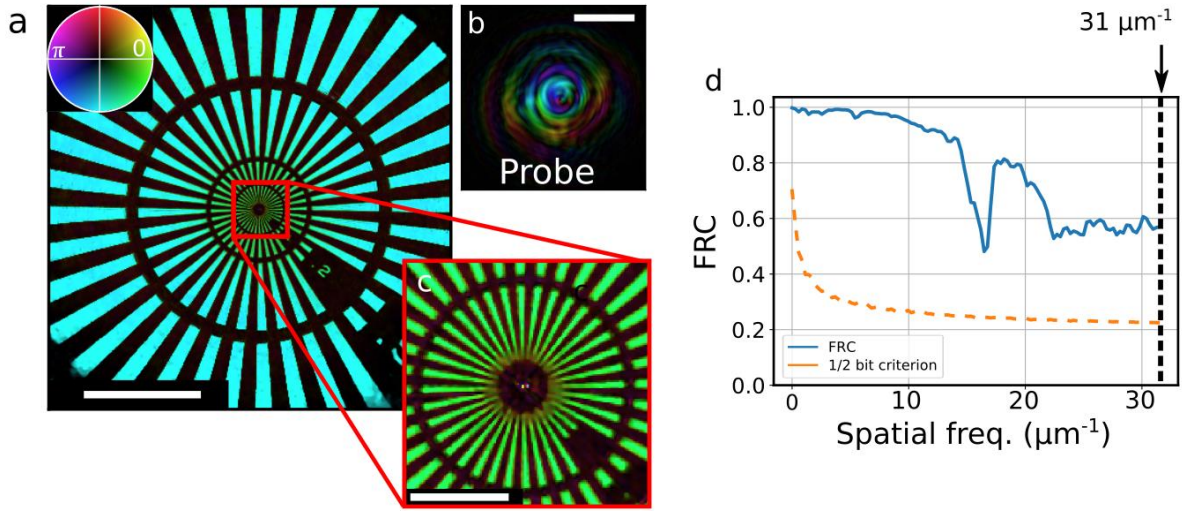

**Figure 3: High-resolution, wide-field of view imaging. a** Reconstructed Siemens star over a field of view of 340 µm$^2$. The corresponding reconstructed illumination (probe) is shown in **b** and features a charge-1 OAM beam. The smallest features are present in the innermost part of the Siemens star, which is shown in **c**. The Fourier ring correlation (FRC) is shown in **d** and indicates a diffraction-limited resolution of 16 nm (31 µm$^{-1}$). The scale bar in **a** and **b** corresponds to 5 µm and the scale bar in **c** corresponds to 1 µm.

*Quantitative, material-specific EUV imaging of an integrated circuit*

Integrated circuits are the foundation of modern computer technology. As the structures of these devices shrink, the characterization of integrated circuits and of the machines used to manufacture them is of great interest. Here, the strengths of ptychography at 92 eV are demonstrated using a highly integrated structure of a conventional solid-state disc (SSD). For this purpose, a lamella of the SSD was extracted and placed in the EUV ptychography microscope. More details on the sample preparation can be found in the methods section.

The reconstructed complex transmission of the lamella is shown in Figure 4 **a**. At the bottom of the lamella, an opaque spot is visible, which can be attributed to contamination and was not visible during the preparation of the sample. In the center, the conducting and insulating structures of the SSD are visible. The reconstructed phase in this area shows a high contrast, which can be explained by numerous materials used for the fabrication and demonstrates the high sensitivity of EUV microscopy. Since the complex object transmission function of the object (i.e. transmissivity and phase shift) retrieved by ptychography contains both quantitative amplitude and phase information, the scattering quotient, averaged along the projection direction, can be accessed. This scattering quotient is independent of the thickness[38,39] and allows identifying different materials by comparison with measured complex refractive indices of different materials[40]. More details on the scattering quotient can be found in the methods section. To ensure a reliable material identification, the transmissivity and phase shift of the object needs to be reconstructed with high precision and referenced to an a priori known area (e.g. vacuum). Although the small, conducting structures of the integrated circuit are resolved in Figure 4 **b**, obvious imaging artifacts are present as well – for instance, the modulations in the vacuum region outside surrounding the specimen. We believe these artifacts can be attributed to small, remaining long-term drifts of the EUV beam on the mask. To reduce the influence of the drifts, we incorporated orthogonal-probe relaxation[23] (OPR) into the forward model of our analysis software[41]. However, because of decoherence effects (more details can be found in the methods section), the forward model in this experiment also required mixed-state (m-s) analysis[42]. We thus implemented a combined OPR and m-s analysis, as detailed in the Methods section. The resulting reconstruction is shown in Figure 4 **c** and shows an improved image quality, as highlighted by the white inset at the right edge of the reconstruction field of view (Figure 4 **d**).

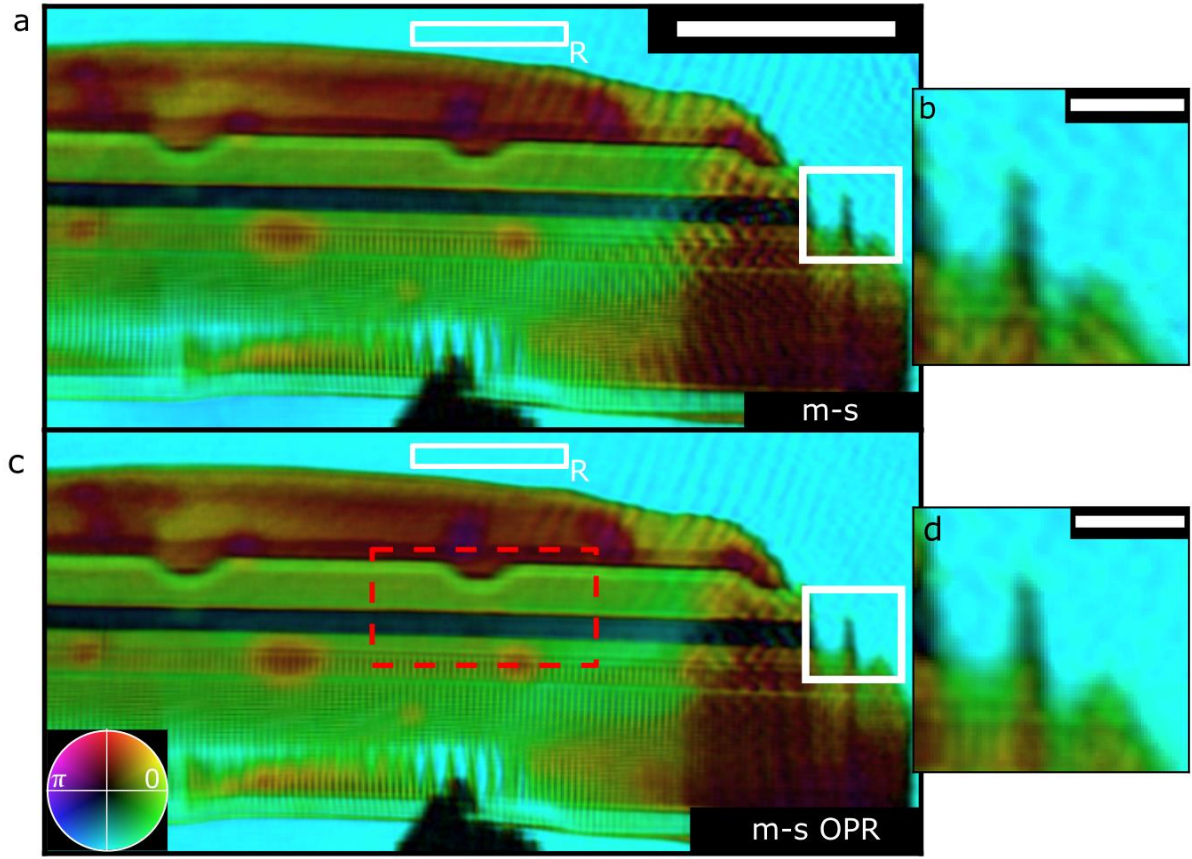

**Figure 4: Quantitative amplitude and phase imaging EUV of an integrated circuit. a** Reconstructed complex transmission of a conventional solid-state disc lamella using the m-s reconstruction model. At the edges of the reconstruction, artifacts are visible, which are particularly evident in the region of interest in **b**. **c** Reconstructed complex transmission using the combined m-s/OPR approach. This results in a reduction of artifacts, as highlighted by the white insets **b** and **d**. The scalebar in **a** corresponds to 5 µm and the scalebar in **b** and **d** corresponds to 1 µm.

The material variety in the integrated circuit can be examined through the scattering quotient. We selected a region of interest (red dashed in Figure 4 **c**), which is shown in Figure 5 **a**. To quantitatively determine the contained materials, a histogram was evaluated for the areas indicated in Figure 5 **a**. Using the m-s method alone (Figure 5 **b**), we encountered difficulties in correctly classifying the contained materials. Using the m-s OPR method (Figure 5 **c**), we obtained well-separated peaks, which can be attributed to silicon nitride (region 1), silicon dioxide (region 2 and 4), and aluminum (region 3). Our analysis was cross-validated by energy-dispersive X-ray spectroscopy (EDX). The nitrogen and oxygen traces are shown in Figure 5 **d**; silicon and aluminum are shown in Figure 5 **e**. The EDX measurements match well with the classification results from our scattering quotient analysis. However, there is one area (region 5) that deviates from the surrounding area. In this region, there appears to be surface contamination with carbon, since the scattering quotient for this area increases ( $f_q(SiO_2) = 2.0$ , $f_q(C) = 5.7$ ). The contamination at this location can be traced back to a high-resolution transmission electron microscopy image, that was acquired after the lamella preparation, which led to a carbon build-up on the surface of the sample. Assuming pure carbon, the thickness of the contamination layer is determined to be 18 nm on each side. The lateral half-pitch resolution of our ptychography reconstruction was determined by an FRC to be 52 nm and is confirmed by resolved features with a half-pitch distance of 78 nm. More details on the estimation of the resolution can be found in the supplement.

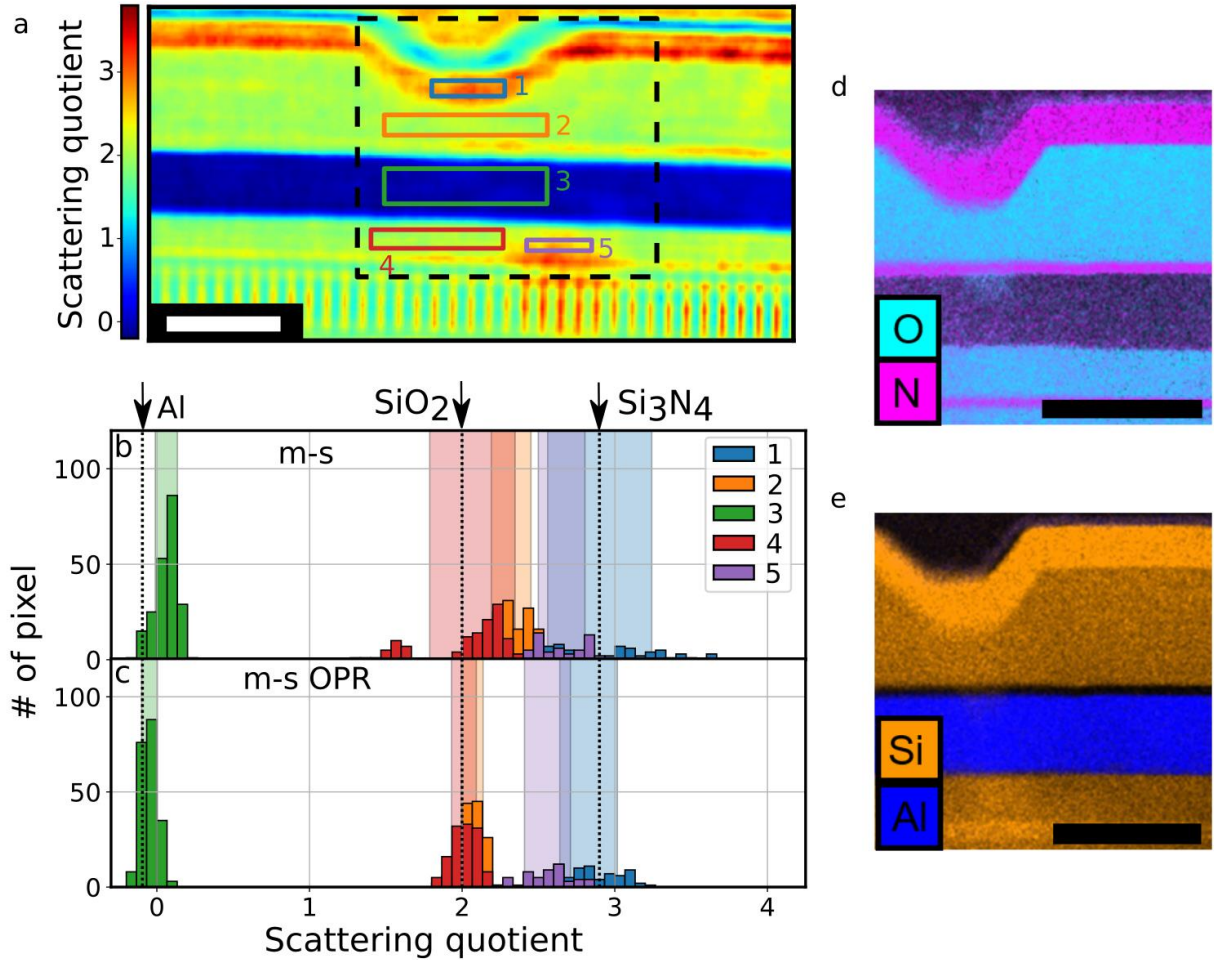

**Figure 5: Material-specific EUV imaging. a** Scattering quotient for a selected region of interest (compare red dashed box in Figure 4 **b**). For each of the numbered regions 1 to 5 a histogram is plotted, as shown in **b** for the m-s reconstruction (Figure 4 **a**) and in **c** for the m-s/OPR reconstruction (Figure 4 **c**). The tabulated scattering quotient[40] for the materials Al (Aluminium, $f_q = -0.1$), SiO$_2$ (silicon dioxide $f_q = 2.0$) and Si$_3$N$_4$ (silicon nitride $f_q = 2.9$) at a photon energy of 92 eV is indicated by a black dotted line in **b** and **c**. The semi-transparent areas indicate plus/minus one standard deviation from the mean scattering quotient in the corresponding areas. Using only the m-s method (**b**), the different materials cannot be identified reliably. For the combined m-s/OPR method (**c**), Al, SiO$_2$, and Si$_3$N$_4$ can be clearly distinguished**.** Energy-dispersive X-ray spectroscopy (EDX) measurements, for a region indicated by a black dashed box in **a,** are shown in **d** for nitrogen and oxygen and in **e** for silicon and aluminum**.** The scale bar in **a**, **d,** and **e** has a size of 1 μm.

*Discussion*

In this paper, we present material-specific high-resolution microscopy in the extreme ultraviolet using a table-top high harmonic light source for the first time. By applying a high photon flux source, structured illumination, and active nanometer-scale stabilization, we achieved a half-pitch lateral resolution of 16 nm on a Siemens star test pattern[43] which is only a factor of 4 larger than the resolution reached by state-of-the-art synchrotron facilities, where a half-pitch resolution down to 3.8 nm has been demonstrated[44]. So far a higher resolution using table-top EUV ptychography has been only reported for a periodic object using a knife-edge test[20].

By combining structured illumination with advanced data analysis, namely a combined mixed-state and orthogonal probe relaxation model, quantitative amplitude and phase imaging with high accuracy is demonstrated. The results were exploited for material-specific imaging of a complex integrated circuit, where a range of different materials have been identified including additional carbon contamination of the sample. This material classification is a novel functionality that significantly expands the capabilities of EUV imaging.

Since the 13.5 nm wavelength features low absorption, samples embedded in micrometer thick silicon environments can be investigated. The large penetration length enables high-resolution imaging and composition analysis of silicon-based nanomaterials, which are important for next-generation rechargeable batteries[3] or metamaterials[4]. Generally, ptychography in the EUV and soft X-ray range will benefit from the presented methods. In principle, the EUV photon energy can be adapted to resonances of other materials (up to 100 eV) by changing the wavelength selecting multilayer mirrors in the setup. For example, the M-edge resonance[45] of iron, nickel, or cobalt can be reached which, in combination with advanced polarization control, will enable imaging of extended magnetic structures (e.g. skyrmions) on a table-top.

Using the presented setup exotic beams with advanced functionalities can be realized, which was demonstrated by the generation of OAM beams of varying charges. This approach offers significantly higher flexibility and lowers experimental complexity compared to state-of-the-art table-top implementations[46–51]. Since OAM beams have already shown a wide variety of advantages in other spectral regions[52] we foresee a plethora of future applications of functional beams in the EUV, including actinic defect inspection via scatterometry with tailored beams[53] and OAM-induced dichroic spectroscopy[54].

**Methods**

*High harmonic generation*

To drive the high harmonic process at 13.5 nm (compare Figure 1), a fiber-based chirped-pulse amplifier operating at a central wavelength of 1030 nm is used. The amplifier consists of four coherently combined fibers that provide 1 mJ pulse energy and 250 fs pulse duration at a repetition rate of 75 kHz, which results in an average power of 75 W. For efficient high harmonic generation, the pulses are compressed by two hollow-core fibers to a pulse duration of < 7 fs with a residual pulse energy of 400 µJ, resulting in an average power of 30 W. The few-cycle pulses are subsequently focused into a gas jet with a backing pressure of 600 mbar of argon, producing a broad EUV continuum reaching up to 12.4 nm wavelength (100 eV). In the next step, the high-power IR laser is separated from the EUV radiation. Four grazing incidence plates[55] separate the broadband EUV continuum from the IR-laser. The remaining IR-light is spectrally filtered by means of two 200 nm zirconium (Zr) foils. Sufficient pointing stability of the EUV beam was achieved by active stabilization of the driving IR laser beam yielding a pointing stability better than 1 µrad at the entrance of the vacuum chamber. The power stability of the EUV beam at 13.5 nm wavelength was characterized to be 0.8 % rms. More details on the general setup of the HHG source can be found in earlier work[56].

*Data acquisition and preprocessing*

All samples were scanned using a Fermat spiral pattern[57]. The distance between adjacent positions was set to 1.0 µm for the Siemens star measurements and 0.7 µm for the integrated circuit. Since the integrated circuit has an elongated geometry, the scan grid was adjusted accordingly. During the measurement, the camera was cooled down to -60°C to minimize thermal noise. The CCD pixels were read-out at a rate of 1 MHz. In the Siemens star measurement, on-chip binning was set to 2x2, while for the integrated structure 1x1 binning was necessary due to a larger sample-mask distance. For each scan position, multiple diffraction patterns at varying exposure times were measured and merged into a single HDR image. The applied stitching method has been described in a previous publication[26]. A table with the most important measurement parameters can be found in the supplementary material. A background correction of the measured diffraction patterns was not necessary, because the vacuum chamber was completely dark. We note that the laser interferometer, employed for position stabilization, operates at a wavelength of 1550 nm at which the CCD (Andor iKon-L) is not sensitive.

*Ptychography reconstruction*

Initial reconstructions directly after the measurements were performed using the difference-map[58] implementation of the GPU-acceleration package of the PtyPy library[59]. The final reconstructions shown here were done using the ptylab framework[41], due to more specialized regulizers (e.g. TV) and models (e.g. mixed-state combined with orthogonal probe relaxation) that are available. For the reconstruction shown here, mPIE[60] combined with the mixed-state method[24] (m-s) is used. The mixed-state method[24] describes the measured far-field intensity $I_j$ at position $j$ as a sum over incoherent modes $P_k$.

$$I_j \sim \sum_k |\mathcal{F}[P_k(r) \cdot O(r - r_j)]|^2$$

Here $O(r)$ corresponds to the complex transmission of the object and $\mathcal{F}[\cdot]$ to the Fourier transform operator modeling far-field diffraction. While HHG sources usually provide a high degree of spatial coherence[16], the m-s modes can be also be employed to mitigate other sources of decoherence such as high-frequency sample vibrations[27], background, detector point-spread, or a finite spectral bandwidth[42]. The reconstructed probes shown in Figure 2 are the dominant illumination mode in each data set, making up more than 50% of the power content. More details on the reconstructed m-s modes can be found in the supplement.

Beam pointing instability and scan stage position drifts that occur on a time scale longer than the exposure time of the camera, cannot be described by the m-s method. Instead, orthogonal probe relaxation (OPR)[23] allows to model probe variations $P_j(r)$ during the scan. This is achieved by relaxing the requirement for a stationary probe along the scan, which can be modeled by a truncated singular value decomposition over the set of probe estimates at each respective scan position $j$,

$$P_j(r) = \sum_i U_i(r) S_i V_{i,j}^*$$

Here $U_i(r)$ corresponds to the $i$th reconstructed eigen-probes, while $S_i V_{i,j}^*$ are expansion coefficients. Thus by allowing the probe modes to be drawn from a larger set of eigen-modes, probe variability can be modeled. We note that mixed states can model decoherence, but require beam stability, while OPR can only model fully coherent probes. For the reconstruction of the integrated circuit (Figure 4 **b**), the m-s method was combined with OPR, which accounts for both probe instability and decoherence effects throughout the measurement,

$$I_j \sim \sum_k \left| \mathcal{F} \left[ \sum_i U_{i,k}(r) S_{i,k} V_{i,j,k}^* \cdot O(r - r_j) \right] \right|^2$$

The hybrid algorithm, which we refer to as *mixed-state orthogonal probe relaxation* (m-s OPR), solves for variable mixed states at each position. In contrast to a previous report[61], where the OPR method was applied to only the first mode of a mixed state model, we applied the OPR method to all mixed states.

Application of the m-s OPR method leads to fewer reconstruction artifacts, rendering image analysis more quantitative - a key to the reliable classification of materials based on the scattering quotient analysis reported here (compare Figure 5). For the quantitative analysis of the integrated circuit, 4 mixed states, each comprising 4 OPR modes, resulted in a model containing a total of 16 probe modes (shown in the supplementary information).

*Material characterization*

Since the complex transmission function of the object reconstructed by ptychography provides both amplitude and phase information, the complex scattering quotient $f_q$ can be calculated[38]. It is defined as the ratio of reconstructed phase $\phi(x, y)$ and the natural logarithm of the reconstructed amplitude $|O(x, y)|$

$$f_q = \frac{\phi(x, y)}{\log(|O(x, y)|)} = \frac{\overline{f_1}}{\overline{f_2}} = \frac{\overline{\delta}}{\overline{\beta}}.$$

Here $\overline{\delta}$ and $\overline{\beta}$ correspond to the averaged refractive index along the projection direction of the measurement, where the refractive index of a material is given by $n = 1 - \delta - i\beta$. The advantage of the scattering quotient is that it is independent of the thickness of the sample. Since ptychography reconstructions are usually invariant under a global phase shift and amplitude scaling, the reconstructed complex transmission must be referenced. In the integrated circuit data, the reconstructed amplitude is referenced to the surrounding vacuum region, which is indicated in Figures 4 **a** and **c** (see white box labeled "R"). Since for the m-s method (Figure 4 **a**) - and to a lower extent also for the m-s OPR method (Figure 4 **c**) - a slight phase curvature in the vacuum region was visible, a low-spatial frequency phase map was interpolated from the surrounding vacuum region and subtracted from the object reconstruction.

*Sample preparation and characterization*

A commercially available SSD-drive (Samsung) has been disassembled and the die of one memory module has been exposed by wet chemical etching. In the next step, a lamella has been excavated by focussed ion beam milling, transferred (*in-situ*) to a carrier structure (Omniprobe-grid) and further thinned to electron transparency using a ZEISS Auriga 40 focussed ion beam workstation.

High-resolution reference analyses of the microstructure and composition were done using a FEI TITAN³ G2 80-300 (S)TEM instrument operated at 300 kV equipped with a high-angle annular dark-field detector (HAADF, Fischione Model 3000) and an energy-dispersive X-ray spectrometry (EDXS) detector (four SDD detectors, FEI company).

*Mask preparation*

A 100x100 µm² $Si_3N_4$ membrane with a thickness of 50 nm was used as a base for the mask fabrication. After coating it with 50 nm of copper (thermal evaporation) from the backside to support charge dissipation during the structuring process, a focused $Ga^+$ ion beam (FEI Helios G3 UC, 30keV, 80pA) was scanned over the Si3N4 surface to etch the desired aperture through the membrane and the copper layer. For this purpose, a black and white bitmap was used to define the structure consisting of 1024x1024 pixels with a pitch of 12 nm by toggling the exposure time between 0 and 200 µs for black and white pixels, respectively. The writing was done within a single pass. Afterwards, additional 220 nm of copper were deposited on the backside to achieve a final absorber thickness of 270 nm Cu + 50 nm $Si_3N_4$. Finally, the aperture shape was confirmed using STEM.

**Data Availability**

The data that support the plots within this paper and other findings of this study are available from the corresponding author upon reasonable request.

**Acknowledgments**

This work was supported by the Federal State of Thuringia (2017 FGR 0076), the European Social Fund (ESF), the Thüringer Aufbaubank (TAB) for funding the junior research group HOROS (FKZ: 2017 FGR 0076), the European Research Council (ERC) under the European Union's

Horizon 2020 research and innovation programm (grant agreement No [835306], SALT) and the Fraunhofer Cluster of Excellence Advanced Photon Sources.

**Author Contributions**

W.E., V.S., L.L., R.K. and A.K. performed the imaging experiments. W.E., L.L and J.R. analyzed the data. L.L. designed the sprial-fresnel zone plate mask. M.S. fabricated the mask and acquired the electron microscope image. H.G. designed the EUV mirror system. M.K. and L.B. prepared the lamella of the integrated structure and performed the EDX and electron microscope measurements. All authors discussed and contributed to the interpretation of the results and the writing of the manuscript. J.R, J.L, T.P, and M.K. supervised the project.

**Competing interests**

The authors declare no competing interests.

# Supplementary information:

## Material-specific high-resolution table-top extreme ultraviolet microscopy

*Wilhelm Eschen, Lars Loetgering, Vittoria Schuster, Robert Klas, Alexander Kirsche, Lutz Berthold, Michael Steinert, Thomas Pertsch, Herbert Gross, Michael Krause, Jens Limpert & Jan Rothhardt*

**Supplementary Note 1: Ptychography setup – optical design**

Here we present optical design considerations of the EUV mirror system regarding spectral filtering properties and optical aberrations. Each of the multilayer mirrors in Figure 1 has a nominal reflectivity of more than 60% at near to normal incidence. Together, the three multilayer mirrors M1-M3 were calculated to yield a spectral FWHM of 0.2 nm. To achieve a small focal spot size, a three-mirror out-of-plane configuration was chosen. The basic idea of the more complicated 3D-geometry of the curved mirror system is to correct the primary astigmatism according to the the well known principle of the Schiefspiegler telescope. The remaining aberrations are coma and second order astigmatism, which are one order of magnitude smaller. Therefore the focus quality can be improved considerably and the number of mirrors remains small to avoid a decreased useable photon flux. Furthermore the use of spherical mirrors guarantees components with high quality surfaces, and avoid expensive toric mirrors, which are fabricated by diamond turning and usually suffer from mid spatial frequency errors[1]. The top and side view schematics of the setup are shown in Supplementary Figure 1. The first mirror (M1) is a planar mirror that steers the beam by 9.9° to the side (around the y axis) and by 4.6° down (around the x-axis). This mirror has no effect on optical performance but has been introduced in order to keep the focused output beam in the same horizontal plane as the input beam. The second (M2) and third mirror (M3) are spherical mirrors. Their focal lengths have been chosen to image the HHG source onto the sample position in the vicinity of the camera flange, while the incidence angles are matched to compensate for astigmatism. M2 has a radius of curvature of 3000 mm and is intended to collimate the divergent EUV beam originating from the HHG source located at a distance of about 1500 mm from M2. M2 steers the beam by 10° around the y-axis; M3 (focal length 300mm) steers the EUV beam by 4.5° around the x-axis and focuses close to the sample.

The resulting EUV focus was characterized by a ptychography measurement and is shown in Supplementary Figure 1 **c**. Since the characterization has been performed 600 µm downstream of the focus, the beam was numerically back-propagated into the focal plane. The resulting complex field is shown in Supplementary Figure 1 **d**. Horizontal and vertical lineouts of the intensity yield a width of 3.8 µm and 2.6 µm (FWHM), respectively.

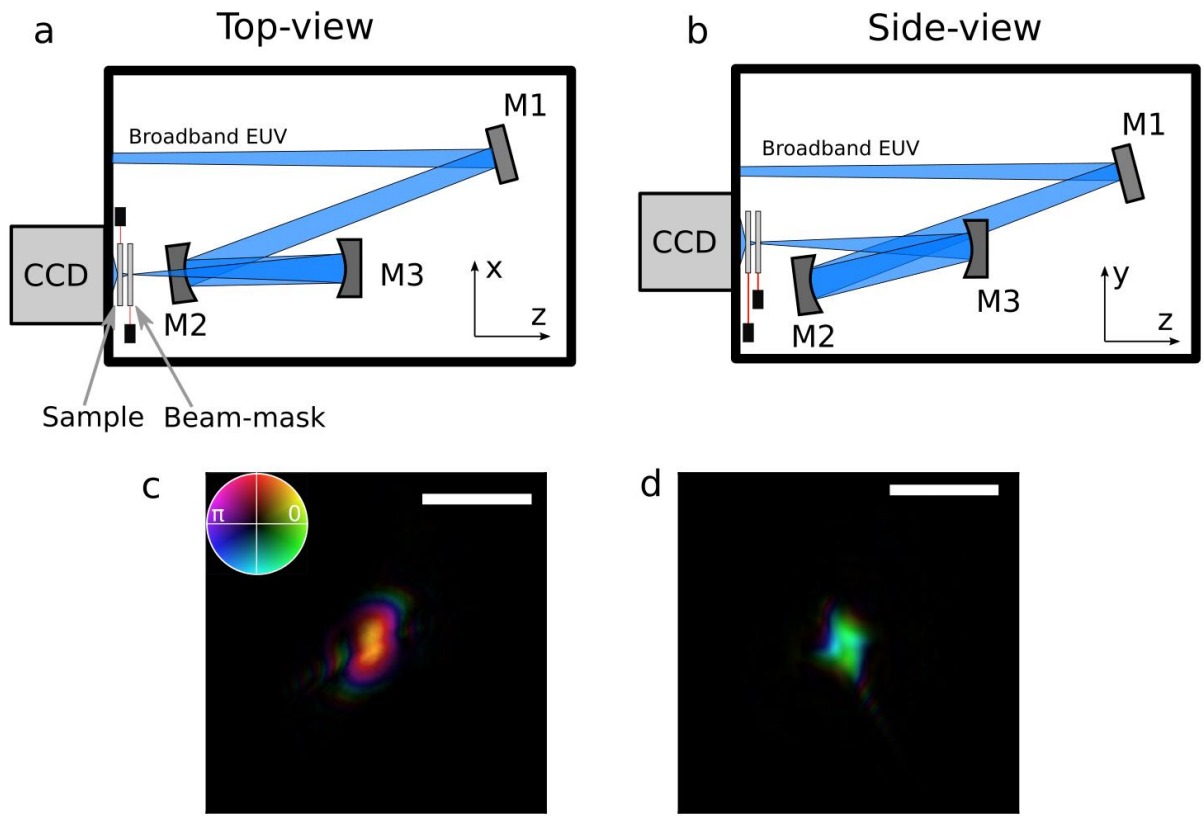

**Supplementary Figure 1: EUV ptychographic imaging setup and characterization of the focal spot**. **a** Top-and **b** side-view of the imaging setup. **c** Reconstructed EUV beam at sample position (600 µm downstream of the focal plane). **d** Reconstructed EUV beam numerically back-propagated into the focal plane. The scale bars in **c** and **d** correspond to 10 µm.

**Supplementary Note 2: Resolution estimation of the integrated structure**

To quantify the achieved lateral resolution of the integrated circuit ptychographic reconstruction (Supplementary Figure 2 **a**), a second identical measurement was performed and the Fourier ring correlation (FRC) was computed from both measurements (Supplementary Figure 2 **d**). The application of the half-bit criterion yields a half-pitch resolution of 52 nm. Supplementary Figures 2 **b** shows a magnified view of the integrated circuit. A lineout along the white line in b is shown in c, indicating features with a half-pitch distance of 78 nm.

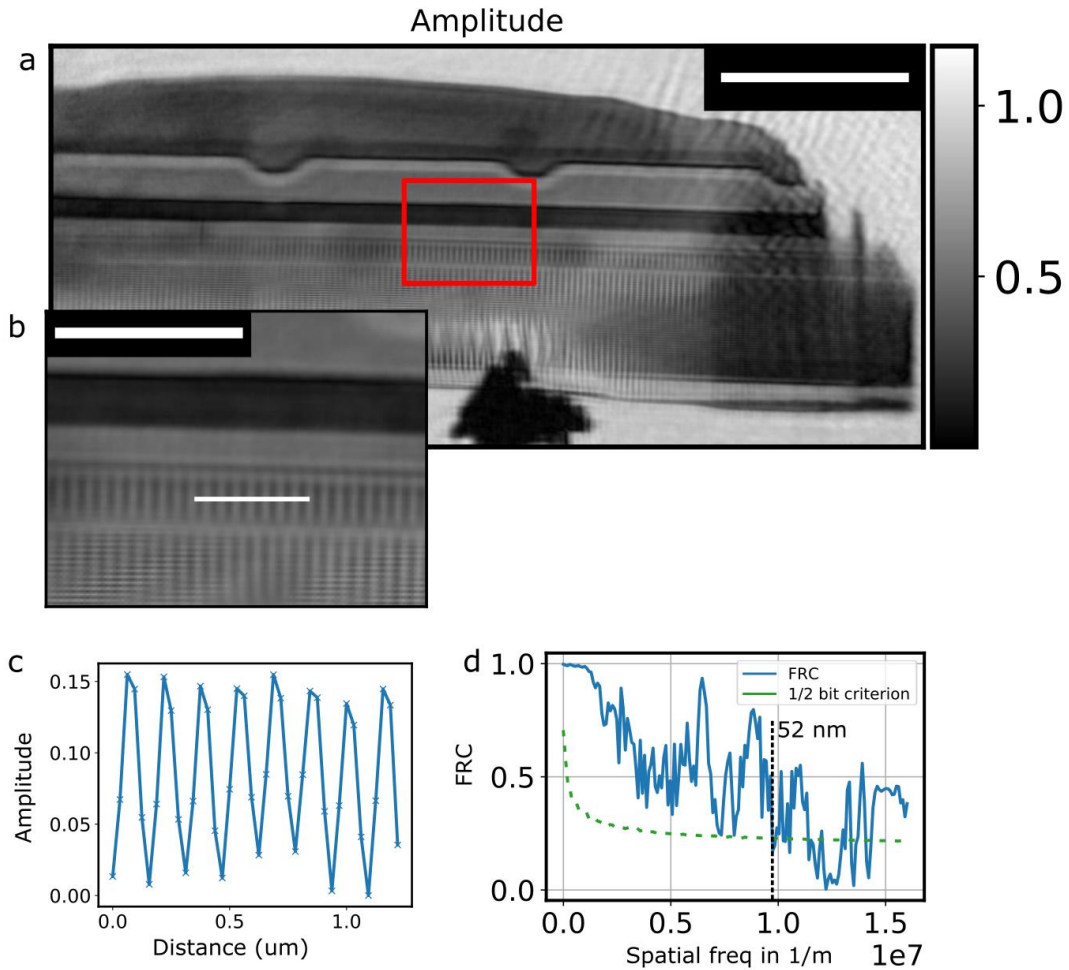

**Supplementary Figure 2:** The reconstructed amplitude of the integrated structure is shown in **a**. **b** magnified view of the red box in **a**. **c** shows a lineout along the white line in **b**, indicating features with a half-pitch distance of 78 nm resolved. The Fourier ring correlation is shown in **d** and results in a half-pitch lateral resolution of 52 nm using the half-bit criterion. The white scale bars in **a** and **b** have a width of 5 µm and 2 µm, respectively.

**Supplementary Note 3: Mixed state ptychography**

For the Siemens star ptychography reconstructions presented in the manuscript, five mutually incoherent probe modes (mixed states) - which henceforth are referred to as *modes* here for brevity – are shown. Supplementary Figure 3 **a** – **e** displays the modes that belong to the data set in Figure 3 of the main text. Each mode was back-propagated into the mask 200 µm upstream of the specimen. The main mode (**a,** Mode 0) contains approximately 60% of the power. Since HHG sources usually provide a high degree of spatial coherence[2], we attribute the m-s modes to other sources of decoherence[3], as mentioned in the Methods section of the main text.

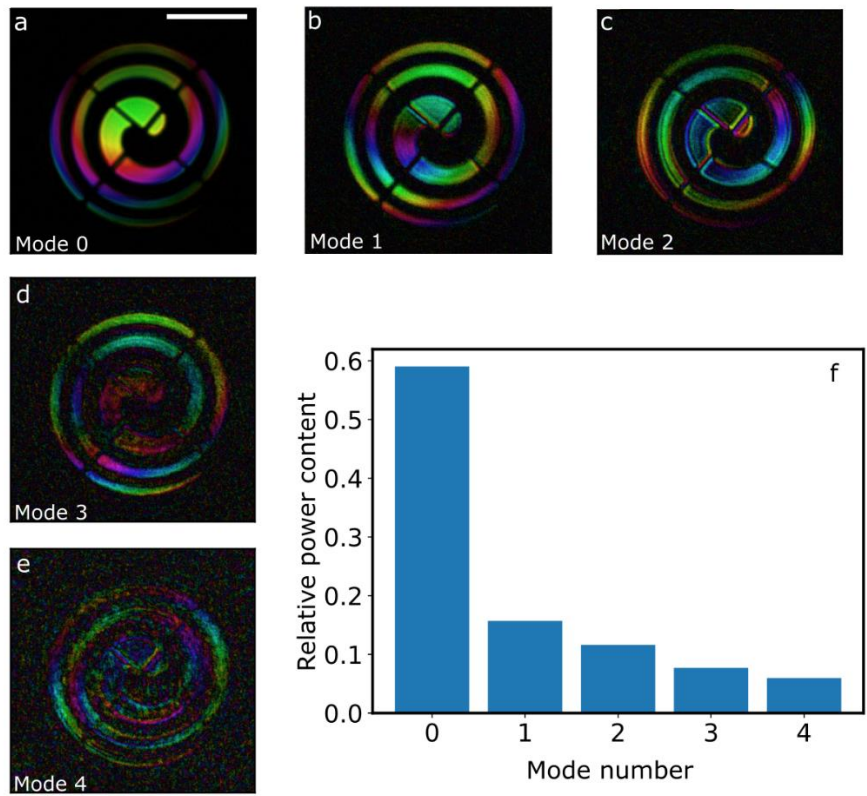

**Supplementary Figure 3:** The reconstructed probe mixed states (modes) from the data set shown in Figure 3 in the main text are shown in **a** - **e**. The relative power content of the modes is shown in **f.** The scale-bar from **a** applies to the plots **a-e** and has a size of 5 µm.

**Supplementary Note 4: Mixed state orthogonal probe relaxation**

The orthogonal probe relaxation (OPR)[4] method allows the reconstruction of probe modes that slightly change during the ptychography scan. Here we applied OPR to all m-s modes in order to maximize the quality of the reconstruction. To keep the memory footprint low, only 4 incoherent modes were used, each of which is synthesized from a linear combination of 4 OPR modes with scan-position-variant expansion coefficients. The resulting m-s OPR modes were back-propagated 550 µm into the mask plane and are shown in Supplementary Figure 4. Here the OPR modes are arranged along the rows and m-s modes (mixed states) are shown along with the columns of the four-by-four matrix of modes. The main mode (m-s mode 0, OPR mode 0) shows the charge-1 OAM mask and a phase curvature which is due to the out-of-focus position of the mask. The OPR modes with power mostly contained in the support region of the mask (e.g. m-s mode 0, OPR mode 2) can be attributed to pointing instability of the EUV beam.

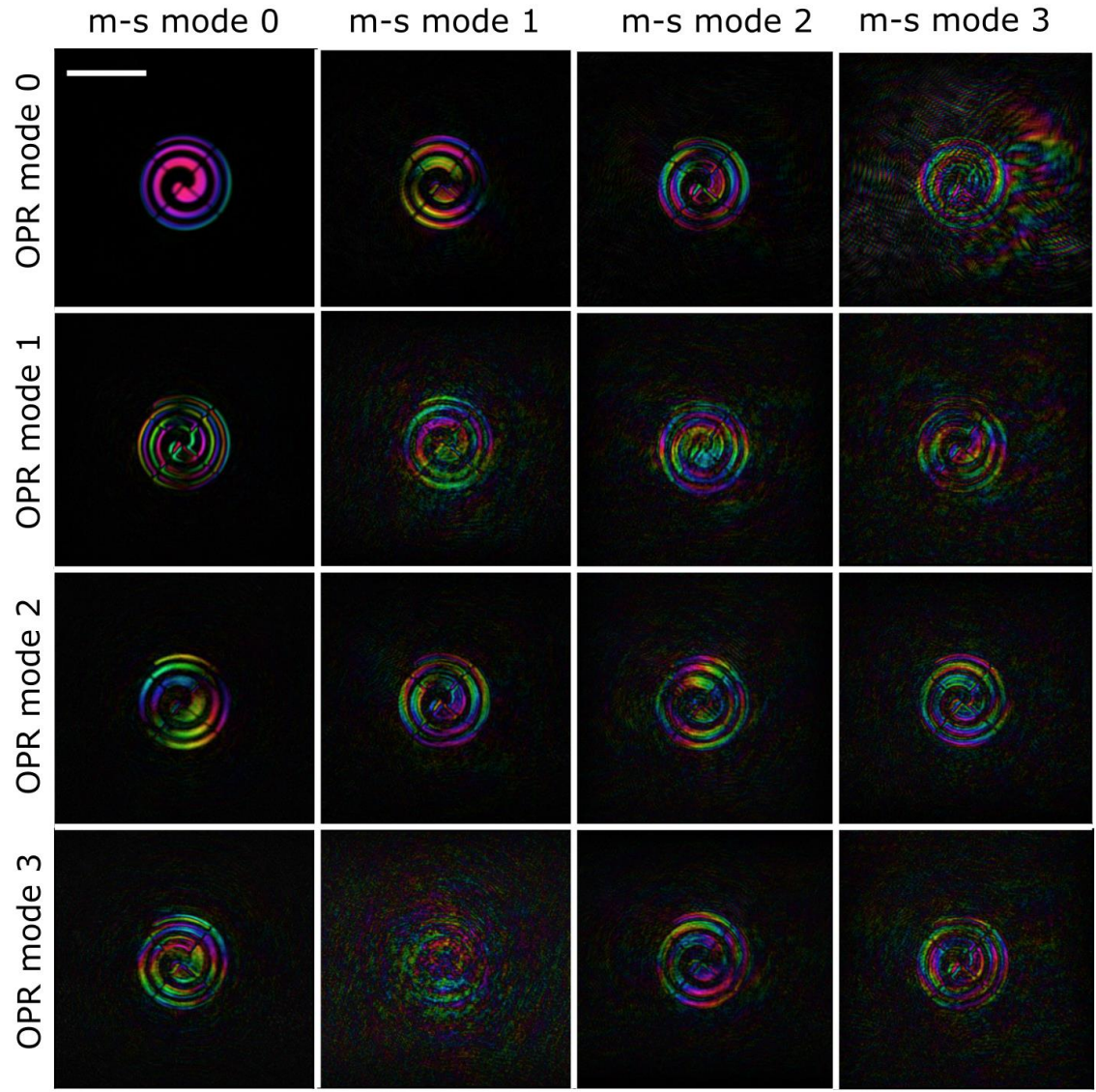

**Supplementary Figure 4:** Overview of jointly retrieved OPR and mixed state modes, arranged along rows and colums. The white scale bar has a size of 10 μm.

The m-s OPR method provides the slowly evolving temporal variation of the m-s modes throughout the scan. Supplementary Figure 5 a shows the m-s modes in the mask plane for the first position (position 0) and last position (position 134). Differences in the m-s mode structure for the first and last position are highlighted by the vertical phase-lineout shown in Supplementary Figure 5 **b** and **c**. The change in the phase can be explained by slow drifts of the EUV-beam during the measurement.

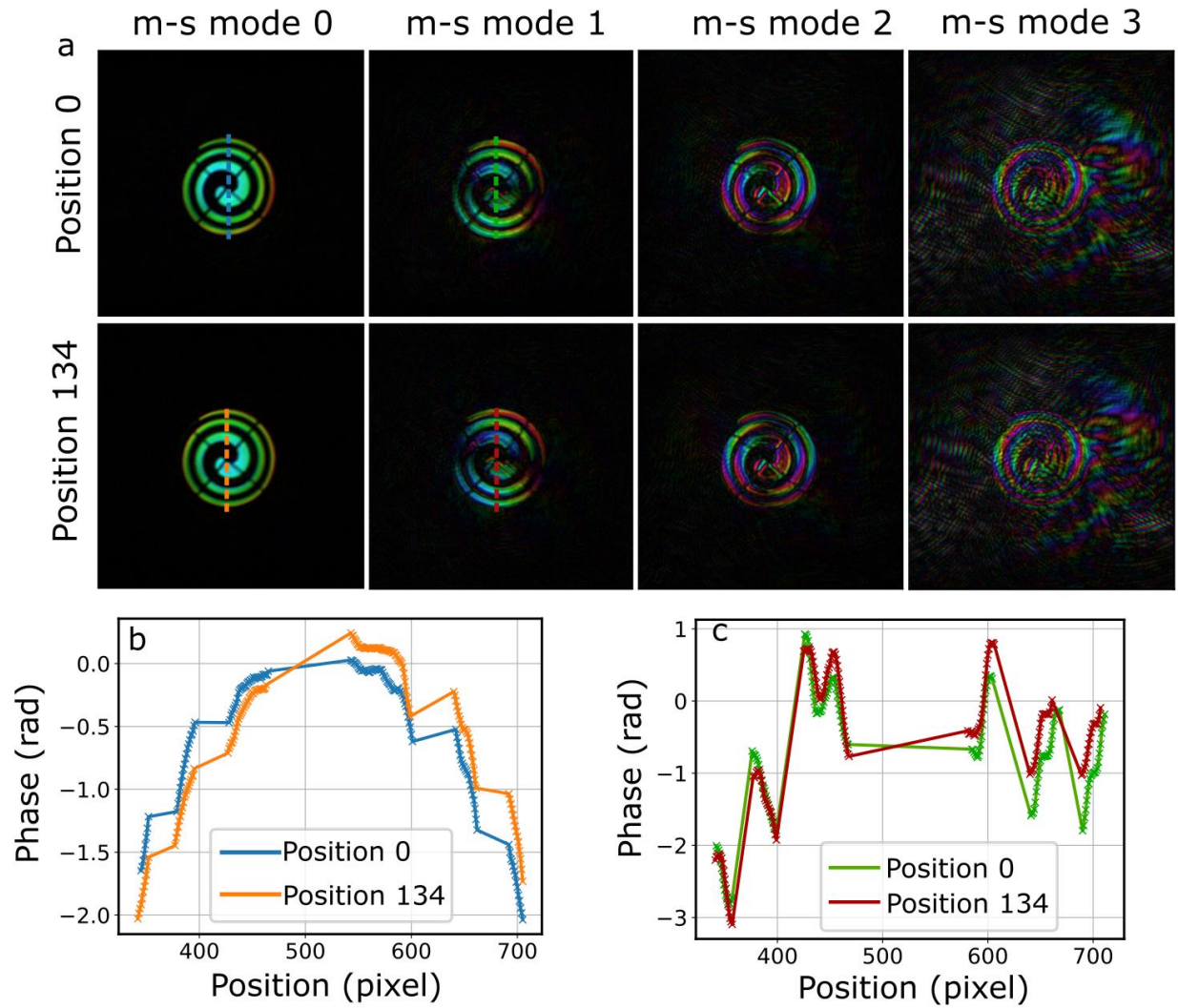

**Supplementary Figure 5: a** shows the reconstructed m-s modes for the first position and the last position of the scan. The vertical phase lineouts for m-s mode 0 and m-s mode 1 are shown in **b** and **c** respectively. It is apparent, that the phase profile of the beam drifted with respect to the mask during the measurement, which can be attributed to a slow drift of the EUV beam pointing.

**Supplementary Note 5: Measurement overview**

| Measurement | # positions | scan step [μm] | on-chip binning | exposure time [s] | mask-sample distance | charge |
|---|---|---|---|---|---|---|
| OAM Fig. 2 | 101 | 1.0 | 2x2 | 5, 30 | 210 μm | 3 |
| Pinhole Fig. 2 | 101 | 1.0 | 2x2 | 2.5, 15 | 170 μm | 0 |
| Figure 3 | 101 | 1.0 | 2x2 | 3, 45 | 200 μm | 1 |
| Integrated circuit | 135 | 0.7 | 1x1 | 0.3, 2.5, 10 | 550 μm | 1 |

**Table 1** Summary of parameters of the ptychography scans reported here. '# positions' – total number of positions of the ptychography scan. 'scan step [μm]' – average distance between adjacent scan positions. 'on-chip binning' – on-chip binning that was used during the measurement. 2x2 on-chip binning results in an effective pixel area of 27 μm x 27 μm. 'exposure time' – CCD exposure times. Multiple exposure times are used for HDR fusion of the diffraction data. 'Mask-sample distance' – distance between the mask and the sample. 'charge' – OAM induced by mask (azimuthal phase shift divided by 2π).